\documentclass[aps,prl,twocolumn,,superscriptaddress,groupedaddress, nofootinbib]{revtex4}  % for review and submission
\usepackage[utf8]{inputenc}
\usepackage{bm}% bold math
\usepackage{color}
\usepackage{hyperref}
\usepackage{amsmath, mathtools}
\usepackage{amssymb}
\usepackage{wasysym}
\usepackage{graphicx} % Add graphics capabilities
\usepackage{booktabs} % ``Proper'' table layout
\usepackage{pdfsync}
\usepackage{verbatim}
\usepackage{latexsym}
\usepackage{dsfont}

\def\sec#1{\section{#1} }

\def\({\left(}
\def\){\right)}
\def\[{\left[}
\def\]{\right]}

\def\a{\alpha}

\def\f#1#2{\frac{#1}{#2}}
\def\g{\gamma}

\def\d{\partial}
\def\de{\delta}

\def\vep{\varepsilon}

\def\l{\lambda}

\def\m{\mu}
\def\n{\nu}
\def\o{\omega}

\def\p{\pi}

\def\s{\sigma}

\def\<{\langle}
\def\>{\rangle}

\providecommand{\abs}[1]{\left\lvert#1\right\rvert}

\usepackage{graphicx}% Include figure files
\usepackage{dcolumn}% Align table columns on decimal point
\usepackage{bm}% bold math

\begin{document}

\preprint{APS/123-QED}

\title{Defect theory of positronium and nontrivial QED relations}% Force line breaks with \\

\author{David M. Jacobs}
\email{djacobs@norwich.edu}
\affiliation{%
Physics Department, Norwich University\\
158 Harmon Dr, Northfield, VT 05663\\% \textbackslash\textbackslash
}%
\affiliation{CERCA, Physics Department\\
Case Western Reserve University\\
Cleveland, OH 44106-7079}%

\date{\today}% It is always \today, today,

\begin{abstract}
An effective theory of the excited states of positronium is derived and some of its consequences are explored.  At large physical separation, the binding of the electron and positron is assumed to be described completely by QED, whereas all short-ranged phenomena, including those within and beyond QED, can be accounted for with energy-dependent quantum defects.  This theory has at least two practical applications. First, it provides an accurate and economical, yet largely QED-independent, means to fit the positronium spectrum in order to predict and compare the outcome of experiments. Second, matching the spectrum in this effective theory to that predicted by QED reveals nontrivial relationships that exist \emph{within} bound-state QED; some higher order contributions to the spectrum may be obtained from lower order contributions. These relations are verified up to order $m\a^{6}$, and predictions are made for the order $m\a^{7}$ and $m\a^{8}$ corrections.  This theory and its extensions to other hydrogenic systems may provide a useful complement to bound-state QED.

\end{abstract}

\pacs{ }% PACS, the Physics and Astronomy
\maketitle

\sec{Introduction}\label{Sec:intro}

The Standard Model of particle physics is a collection of quantum field theories that describe the fundamental interactions of matter. It includes quantum electrodynamics (QED), the theory describing the interactions of electrically charged particles with photons. Agreement between QED and some experimental measurements has reached a precision of approximately one part in a billion, making QED one of the most accurate physical theories ever devised \cite{Hanneke:2008tm,Morel:2020dww}.

The precision with which QED has been used to describe positronium (Ps), the bound state of an electron and positron, is finally being matched by advancements in laser spectroscopy made within the past few decades. There are now efforts to create long-lived Rydberg states of Ps, creation of which are being used for precision spectroscopic measurements \cite{Cassidy:2018tgq}. Interestingly, and pertinent to this article, is a recent measurement of a Ps transition frequency that disagrees with the QED prediction at the level of $4.5\sigma$ \cite{Gurung:2020hms}. While such a discrepancy between experiment and theory could turn out to be an experimental or theoretical error, it may be a sign of new physics \emph{beyond} the Standard Model. This is exciting because BSM physics is expected to be discovered eventually, as the Standard Model itself is likely insufficient to explain several phenomena, such as neutrino oscillations \cite{Mohapatra:1998rq}.% or the existence of dark matter \cite{Bertone:2004pz}.

Whether or not the results of \cite{Gurung:2020hms} turn out be a manifestation of a BSM effect, it would be useful to have a robust theory that describes positronium without reference to \emph{any} particular fundamental model, but a theory that is predictive regardless of the ultimate microscopic nature of reality.  In this article, an effective theory of Ps is presented; in essence, it is a relativistic, two-body quantum defect theory\footnote{Quantum defect theory traces its history to descriptions of the (nonrelativistic) Rydberg states of large alkali atoms \cite{Seaton_1983}.}. Although it is not a field theoretic approach, its remarkable predictive power reveals nontrivial relations that must exist within any approach to bound state QED, at least insofar as positronium is concerned. 

Within perturbative QED, the lowest-order computation of the energy levels of positronium gives 
\begin{equation}\label{lowest_Ps}
E_n =-\f{m\a^2}{4n^2}\,,
\end{equation}
where $m$ is the electron mass, $\alpha$ is the fine structure constant, and $n$ is a positive integer. Increased theoretical accuracy accounting for relativistic effects (kinetic, spin coupling, and radiative) is achieved with corrections to \eqref{lowest_Ps} that have increasingly large exponents of $\a$; for example, corrections at the next order of precision have a characteristic size $m\a^4$. Further corrections require the field-theoretic machinery of QED, using either the Bethe-Salpeter formalism \cite{Salpeter:1951sz} or the more widely used nonrelativistic QED (NRQED) \cite{Caswell:1985ui}. All corrections to \eqref{lowest_Ps} up to order $m\alpha^6$ have been computed as of the late 1990’s (see, e.g., \cite{Pachucki:1997vm} and \cite{Czarnecki:1999mw}). Some of the order $m\alpha^7$ corrections have been computed, but the project of computing the full positronium spectrum up to order $m\alpha^7$ -- a cumbersome task involving the computation of many Feynman diagrams -- is still on-going \cite{Adkins:2019zgu}.

Here we take an approach influenced by \cite{BeckThesis}, and then expanded upon in \cite{Jacobs:2015han} and \cite{Jacobs:2019woc} in which quantum mechanics is used to provide an effective, long-distance description of bound systems. The main feature used in \cite{Jacobs:2015han} and \cite{Jacobs:2019woc} is a so-called \emph{boundary of ignorance} that explicitly excludes short-ranged interactions; in this sense it can be thought of as a real-space UV cutoff.  Instead of modeling those interactions, energy-dependent boundary conditions on the system's wavefunction encode information about them.  These boundary conditions therefore parametrize an effective quantum mechanics similarly to the way the effective Lagrangian is used in field theory to parametrize the effects of high energy degrees of freedom that have been integrated out\footnote{See \cite{Burgess:2016lal} for a complementary field theoretic approach to the one described in \cite{Jacobs:2015han} and \cite{Jacobs:2019woc}.}.
In the context of a hydrogenic system, these boundary conditions were shown in \cite{Jacobs:2019woc} to manifest themselves in energy dependent quantum defects, a key result that will be applied below.

\sec{Long-range Effective Potential}\label{Sec:Effective_Potential}

To arrive at a relativistic effective theory for positronium, we begin by outlining the desired form of the effective Hamiltonian, namely one that describes the long distance interaction of two charged particles of identical mass. In the center-of-momentum frame the time-independent Schroedinger equation is presumed to take the form
\begin{equation}\label{SchrodingerEqn}
\(2\hat{\vep}-2m + U_\text{eff}(r)\)\psi=E\psi
\end{equation}
where $r$ is the relative separation of the two particles, $\vec{p}\equiv\vec{p}_1=-\vec{p}_2$ is the conjugate momentum operator, and $E$ is the energy of the system less the two masses.  The energy operator
\begin{equation}
\hat{\vep}=\sqrt{\vec{p}^{\,2}+m^2}\,,
\end{equation}
would be implemented in practice by expanding to any desired order in the quantity $\vec{p}^{\,2}/m^2$. The long-distance effective potential, $U_\text{eff}(r)$ could in principle depend on quantum numbers, such as angular momentum. However, as shown below, such an explicit dependence is absent because interactions that depend on angular momentum are short-ranged, falling off with distance faster than $r^{-1}$.

Assuming that QED sufficiently describes this system at long distance, we obtain $U_\text{eff}(r)$ by matching the calculation of the elastic differential scattering cross section of the two particles using two different methods, one  field theoretic and one quantum mechanical. Following \cite{berestetskii1982quantum}, the \emph{field theoretic} calculation in the center-of-momentum frame gives
\begin{equation}\label{diffcross_QFT}
\f{d\s}{d\Omega}= \f{1}{256\p^2 \vep^2} \abs{M_{fi}}^2 \,,
\end{equation}
where $\vep$ is the relativistic energy of either particle, individually\footnote{This is a difference in notation from \cite{berestetskii1982quantum}, wherein $\vep$ refers to the total energy of the two particles.}. We compute the scattering amplitude, $M_{fi}$ using QED for two distinguishable spin-1/2 particles with electric charges $e_1$ and $e_2$ that scatter elastically and at very long distance. In other words, the momentum transfer, $\vec{q}$ is much smaller than any natural momentum or energy scale in the system; therefore it should satisfy
\begin{equation}
\abs{\vec{q}}\equiv\abs{\vec{p}\,'-\vec{p}}\ll m \a^2 \,.
\end{equation}
The lowest order amplitude, due to a single photon exchange, is\footnote{One could also consider the virtual annihilation process; however, because $\abs{\vec{q}} \ll m\a^2 \ll m$, we neglect this. In any case, such processes are accounted for by a complex quantum defect.}
\begin{equation}
%M_{fi}=-e^2\bar{u}(p')\g^\m u(p)D_{\m \n}(q)\bar{u}(-p)\g^\n u(-p')\,,
M_{fi}=e_1e_2\(\bar{u}'\g^\m u\)D_{\m \n}(q)\(\bar{u}'\g^\n u\)\,,
\end{equation}
where $u=u(\vec{p},m)$ and $\bar{u}'=\bar{u}(\vec{p}\,',m)$. The photon propagator in the Feynman gauge is
\begin{equation}
D_{\m \n}(q)=\f{4\p}{q^{2}} g_{\m\n}\,,
\end{equation}
where $q^{2}=q_0^2-\vec{q}^{\,2}$ is the square of the virtual photon four-momentum; however, $q_0=0$ identically because the masses of the two particles are equal.
 As the $\vec{q}\to 0$ limit is taken, 
 \begin{equation}\label{spin_series_form}
\bar{u}'\g^\m u   = 2p^\m + {\cal O}\(\abs{\vec{q}}\)\,, %\bar{u}_1\g^\m u_1 
\end{equation}
and it follows that
\begin{equation}\label{qtozero_matrix_element}
\lim_{\vec{q}\to0}M_{fi} = -\f{16\pi e_1 e_2}{\vec{q}^{\,2}} \( \vep^2 +\vec{p}^{\,2} \)\,.%\(1+ {\cal O}\(\f{\abs{\vec{q}}}{m_{1,2}}\)\)
\end{equation}
The omitted terms in \eqref{qtozero_matrix_element} include, in the nonrelativistic limit, those of order $\f{\abs{\vec{q}}}{m}$, consisting of the well-known spin-orbit and spin-spin coupling terms, and other short-ranged corrections that appear in the real-space Breit potential. Such terms are omitted consistently with the fact that they correct $U_\text{eff}$ with terms that fall off with distance\footnote{In fact, all of these corrections in the Breit potential scale with distance as either $r^{-3}$ or are proportional to a delta function centered about the origin \cite{berestetskii1982quantum}.} faster than $r^{-1}$, the scaling of the leading term in the potential. Additional loop diagrams could be drawn and computed, in principle; however, such diagrams provide corrections to \eqref{qtozero_matrix_element} that are subdominant in the $\vec{q}\to0$ limit (see, e.g., \cite{Peskin:1995ev} and \cite{HelayelNeto:1999ut}) so they are also omitted.

Dropping the terms mentioned above has a very important simplifying consequence as we attempt to use the \emph{quantum mechanical} method to compute the differential cross section. To show this, we first consider that equation \eqref{qtozero_matrix_element} indicates, by a Fourier transform, that $U_\text{eff}(r) \propto r^{-1}$ at leading order. Second, we will show that \eqref{SchrodingerEqn} can be manipulated to take the asymptotic form 
\begin{equation}\label{psqr_posit}
\vec{p}^{\,2} \psi \sim \(C_0+\f{C_1}{r} \)\psi\,,
\end{equation}
for some constants $C_{0}$ and $C_{1}$. It follows that we may utilize the relation
\begin{equation}\label{commutation_simplification}
\[\vec{p}^{\,2},U_\text{eff}(r)\]\psi=0~~~~\mbox{(effectively)}\,,
%\[\del^2,\f{1}{r}\]=0~~~~\mbox{(effectively)}\,,
\end{equation}
because the terms neglected on the right hand side of \eqref{commutation_simplification} scale as either $r^{-2} \d_r\psi$ or a delta function centered about the origin, multiplied by $\psi$; such terms are subdominant at long distance and therefore may be discarded.

We now calculate the differential scattering cross section using equation \eqref{SchrodingerEqn}. Take the two free particles to have incoming momenta $\vec{k}$ and $-\vec{k}$, respectively, so that their relativistic energies are
\begin{equation}
\vep\equiv\sqrt{\vec{k}^{\,2}+m^2}\,,
\end{equation}
and therefore
\begin{eqnarray}
E=2\vep- 2m\,.
\end{eqnarray}
Equation \eqref{commutation_simplification} allows us to use purely algebraic means to show that \eqref{SchrodingerEqn} leads to
\begin{equation}\label{Sch_scatt}
\vec{p}^{\,2} \psi = \(\vec{k}^{\,2}  - \vep U_\text{eff}(r)  + {\cal O}\(U_\text{eff}^2\)\)\psi\,,
\end{equation}
which is consistent with  \eqref{psqr_posit} upon dropping the ${\cal O}\(U_\text{eff}^2\)$ terms, which fall off faster than $r^{-1}$. A first order (Born) scattering analysis may then be used to show that
\begin{equation}\label{diffcross_Schrodinger}
\f{d\s}{d\Omega}=  \f{1}{16\p^2}\vep^2 \abs{\tilde{U}_\text{eff}(\vec{q})}^2\,,
\end{equation}
where $\tilde{U}_\text{eff}(\vec{q})$ is the Fourier transform of $U_\text{eff}(r)$.

Finally, by using \eqref{qtozero_matrix_element} to match  \eqref{diffcross_QFT} with \eqref{diffcross_Schrodinger}, we find the  effective potential in Fourier space to be %given by its Fourier space form,
\begin{equation}\label{U_gen_q-space}
%\tilde{U}(\vec{q},\vec{p})
\tilde{U}_\text{eff}(\vec{q})=\f{4\pi e_1e_2}{\vec{q}^{\,2}}\(1+\f{\vec{p}^{\,2}}{\vep^2}   \)\,,
%\tilde{U}(\vec{q},\vec{p})=\f{4\pi e_1 e_2}{\vec{q}^{\,2}}\(1+\f{\vec{p}^{\,2}}{\vep_1 \vep_2}  - \f{1}{\vep_1 \vep_2}\f{\(\vec{q}\cdot \vec{p}\)^2}{q^2} \)
\end{equation}
where the sign is determined by the physical requirement that the potential be repulsive (attractive) for like (opposite) charges.  It is not obvious that this analysis can be trusted to arbitrarily high $\vec{p}$ because the creation of additional particles is not accounted for here. On the other hand, because the virtual photon has vanishing momentum in this long-distance limit, it would not seem that additional out-going massive particle states could be populated.

Although this derivation proceeded for the interaction of two spin-1/2 particles, the fact that it is insensitive to spin effects -- a manifestation of dropping corrections to  \eqref{spin_series_form} that are higher order in $\vec{q}$   -- suggests that it is equally applicable to spin-0 particles, and likely applies to fermions and bosons, generally.

\sec{Effective Theory of Positronium}\label{Sec:Long-range_solutions}
When the two particles involved are an electron and positron, we set the charges $e_1=-e=-e_2$, and let $m$ be the mass of the electron. The real-space potential is found by Fourier transform of \eqref{U_gen_q-space}:
\begin{equation}\label{final_Ps_effective_potential}
U_\text{eff}(r)=-\f{\a}{r}\(1+\f{\vec{p}^{\,2}}{\vep^2} \)\,,
\end{equation}
where the fine structure constant, $\a=e^2$. In the $\abs{\vec{p}}\ll m$ limit \eqref{final_Ps_effective_potential} is just the Coulomb potential. In the ultra relativistic limit, $\abs{\vec{p}}\gg m$, the potential is enhanced by a factor of two, a feature reminiscent of the well-known factor-of-two difference between some Newtonian and general relativistic predictions.

Because it contains the momentum operator, it would appear that $U_\text{eff}(r)$ should itself be treated as an operator; however, the analysis is again simplified by the use of \eqref{commutation_simplification}. The time-independent Schrodinger equation \eqref{SchrodingerEqn} using the potential given in \eqref{final_Ps_effective_potential} may be solved algebraically in this long-distance limit to be of the form indicated in \eqref{psqr_posit}. We consistently discard all terms that fall off with distance faster than $r^{-2}$, and find the relativistic radial Schrodinger equation to take the same form as its nonrelativistic counterpart, namely 

\begin{equation}\label{radial_sch_eq}
R''(r)+\f{2}{r}R'(r) -  \f{\tilde{\ell}\(\tilde{\ell}+1\)}{r^2}R(r) + \f{m\tilde{\a}}{r}R(r)=q^2 R(r)\,,
\end{equation}
where
\begin{equation}\label{tilde_ell_eqn}
\tilde{\ell}\(\tilde{\ell}+1\)\equiv \ell\(\ell+1\) + {\cal O}\(\a^2\)\,,
\end{equation}

\begin{equation}\label{tildealpha_eqn}
%\tilde{\a} = \a\(1 + \f{3E}{2m} - \f{E^2}{4m^2}  +  \f{E^3}{8m^3} - \f{E^4}{16m^4} + {\cal O}\(\f{E}{m}\)^5\)
\tilde{\a} = \a\(1 + \f{E}{m} + \f{E}{E+2m}\)\,,
\end{equation}
%\flag{\begin{equation}
%%\tilde{\a} = \a\(1 + \f{3E}{2m} - \f{E^2}{4m^2}  +  \f{E^3}{8m^3} - \f{E^4}{16m^4} + {\cal O}\(\f{E}{m}\)^5\)
%\tilde{\a} = \a\(1 + \f{3E+\f{E^2}{m}}{E+2m}\)
%\end{equation}}
and
\begin{equation}\label{qsqr def}
%q^2=-mE\(1+\f{E}{4m} +\dots \)  %{\cal O}\(\f{E^3}{m}\)
q^2=-mE -\f{E^2}{4}\,.  %{\cal O}\(\f{E^3}{m}\)
\end{equation}
Note that we do not display the ${\cal O}\(\a^2\)$ terms in  \eqref{tilde_ell_eqn} because determining them consistently would have required keeping the  ${\cal O}\(U_\text{eff}^2\)$ terms that were discarded in equation \eqref{Sch_scatt}. Following the analysis in \cite{Jacobs:2015han} and \cite{Jacobs:2019woc}, the solution to \eqref{radial_sch_eq} that is generally normalizable is %decays sufficiently rapidly with growing $r$ %\cite{Jacobs:2021cdb}
\begin{equation}
R(r)= e^{-qr}r^{\tilde{\ell}}\,{\cal U}\[1 + \tilde{\ell} - \f{m\tilde{\a}}{2q},2 + 2\tilde{\ell},2qr\]\,,
\end{equation}
where ${\cal U}$ is the Tricomi hypergeometric function. 

The canonical solutions\footnote{These special solutions are those that are regular at the origin.} are recovered when the quantity $1 + \ell - \f{m\a}{2q}$ is equal to a nonpositive integer, conventionally defined by
\begin{equation}
q=\f{m \a}{2n}\,,
%q=\f{m\tilde{\a}}{2n)}\,,
%q=\f{m\tilde{\a}}{2\(n+\tilde{\ell}\)}\,,
\end{equation}
where the principle quantum number $n\geq \ell+1$, and there exist energy degeneracies amongst states of different orbital angular momentum, $\ell$. %where $n=1,2,3\dots$. It follows that

As for any other atomic system, the real positronium wavefunctions do not follow their canonical solutions exactly. To account for this we make the definition
\begin{equation}\label{q_ansatz}
%q\equiv\f{m \a}{2\n}\,,
q\equiv\f{m\tilde{\a}}{2\n}\,,
%q=\f{m\tilde{\a}}{2\(n+\tilde{\ell}\)}\,,
\end{equation}
%where $n=1,2,3\dots$. It follows that
where $\n$ is the effective quantum number, and introduce the quantum defect in the standard way \cite{Seaton_1983} by letting
\begin{equation}\label{nu_ansatz}
\n=n-\de_{\ell s j}(E)\,.
\end{equation}
It must be remembered that $\n$ depends implicitly on the bound state energy $E$ and on the orbital, total spin, and total angular momentum quantum numbers, $\ell$, $s$, and $j$, respectively; we suppress this notation wherever possible for clarity. As first explored in \cite{BeckThesis} and later expanded upon in \cite{Jacobs:2019woc}, the quantum defect $\de_{\ell s j}(E)$ captures the effect of short-ranged interactions and, therefore, the deviations from the canonical solutions. Similar to \cite{Jacobs:2019woc}, we posit the defects to have the analytic form %\cite{Jacobs:2021cdb}
%\footnote{In \cite{Jacobs:2019woc} it is posited shown that the defect take the form of a series expansion in $q^2$, but a series in $E$ is equivalent here, given \eqref{qsqr def}. }
\begin{equation}\label{delta_general_ansatz}
\de_{\ell s j}(E) = \de_{\ell s j(0)}+\l_{\ell s j(1)} \f{E}{\Lambda} + \l_{\ell s j(2)} \(\f{E}{\Lambda}\)^2  +\dots  \,,
%\de_{j\ell s}(E) = \de_{j\ell s,0}+\f{\de_{j\ell s,2}}{\(n-\de_{j\ell s,0}\)^2} + \dots
\end{equation}
where $\Lambda$ is a high energy scale and the free parameters $\de_{\ell s j(0)}$, $\l_{\ell s j(1)}$, etc. must be fit to experimental data or a UV-complete model, such as QED.

Given equation \eqref{tildealpha_eqn}, \eqref{qsqr def} and \eqref{q_ansatz}, the bound state energies of Ps apparently obey
\begin{equation}\label{the_main_event}
E=m\(\sqrt{2+\f{2\n}{\sqrt{\(\n^2+\a^2\)}}}-2\)\,.
\end{equation}
Equation \eqref{the_main_event} -- using \eqref{nu_ansatz} and the defect ansatz \eqref{delta_general_ansatz} -- is the central result of this work. It shares similarity with the results of  \cite{Johnson_1979}, wherein a relativistic quantum defect theory was derived for the Dirac-Coulomb problem, i.e., for a relativistic electron bound to an infinitely massive atomic core.  However, equation \eqref{the_main_event} is remarkable in that it apparently accounts for recoil effects.
To demonstrate that equation \eqref{the_main_event} is plausible, consider the hypothetical limit in which $\de_{\ell s j}(E)\to0$. In that case, $\n\to n$, and a series expansion in small $\alpha/n$ may be performed, indicating %, corresponding to the scenario in which there are \emph{no} short-ranged physical effects of relevance
\begin{equation}\label{Ps_energy1}
E=-\f{1}{4}\f{m\a^2}{n^2} +\f{11}{64}\f{m\a^4}{n^4} - \f{69}{512}\f{m\a^6}{n^6}+{\cal O}\(\f{\a^8}{n^8}\)%\\
% +\f{1843}{16\,384}\f{m\a^8}{n^8}  - \f{12\,767}{131\,072}\f{m\a^{10}}{n^{10}}+ \dots %
\end{equation}
%then
The first three terms are already known to be common to all positronium levels according to QED; see, e.g.,  \cite{Czarnecki:1999mw} or \cite{Zatorski}.  One might say that, in this hypothetical limit, all terms in \eqref{Ps_energy1} beyond the first represent \emph{kinetic} relativistic corrections, as opposed to relativistic corrections that are short-ranged in nature, such as those involving spin coupling or radiative effects.

Lastly, we must account for the fact that Ps is an unstable system with decays that occur when its constituents annihilate. To account for the decaying time evolution of these unstable states, the energies are treated as generally complex, i.e.
\begin{equation}
E=\o -i \f{\Gamma}{2}\,,
\end{equation}
where $\o$ is the real part of the energy -- measurable via spectroscopy -- and $\Gamma$ is the decay rate of the state, assumed only to result from annihilations\footnote{In other words, $\Gamma$ does not capture radiative transitions \emph{bewteen} states of Ps.}. It is straightforward to show that
\begin{equation}
\o=-\f{m\a^2}{4n^2}-\f{m\a^2}{2n^3} \de_{(0),\text{re}}   + \dots\,, %+\f{11}{64}\f{m\a^4}{n^4}
\end{equation}
whereas
\begin{equation}
\Gamma= \f{m\a^2 \de_{(0),\text{im}}}{n^3} +\dots\,,%+\f{3m\a^2}{n^4} \de_{(0),\text{im}} \de_{(0),\text{re}} - \f{11 m\a^4 \l_{(1),\text{im}}}{8n^5} + \dots
\end{equation}
where the omitted terms are higher order in the small quantity $\a/n$ and/or the defect parameters.

\sec{Robustness of the theory}\label{Sec:Fit_to_QED}

\begin{figure}[htp]
  \begin{center}
    \includegraphics[width=8.6cm]{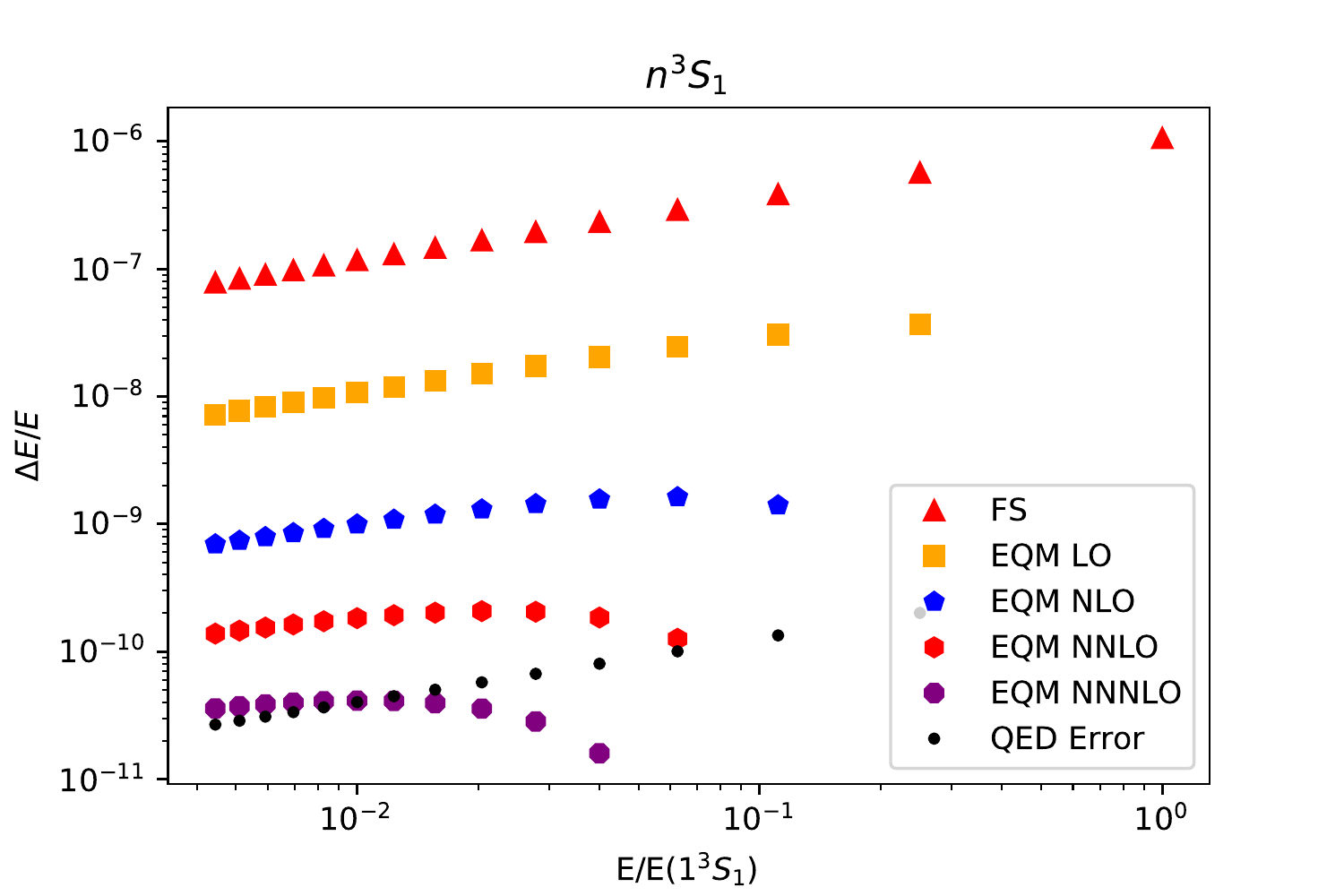}
  \end{center}
  \caption{Relative errors in the predicted $n^3S_1$ state energies, using the ${\cal O}\(m\a^6\)$ QED calculations as a fiducial model. The energies indicated on the horizontal axis are normalized to the lowest energy, $E(1\,^3S_1)$. For reference, the ${\cal O}\(m\a^7\)$ QED error estimate from \cite{Czarnecki:1999mw} is indicated with black dots.}
  \label{3S1_plot}
  \end{figure}

\begin{figure}[htp]
  \begin{center}
    \includegraphics[width=8.6cm]{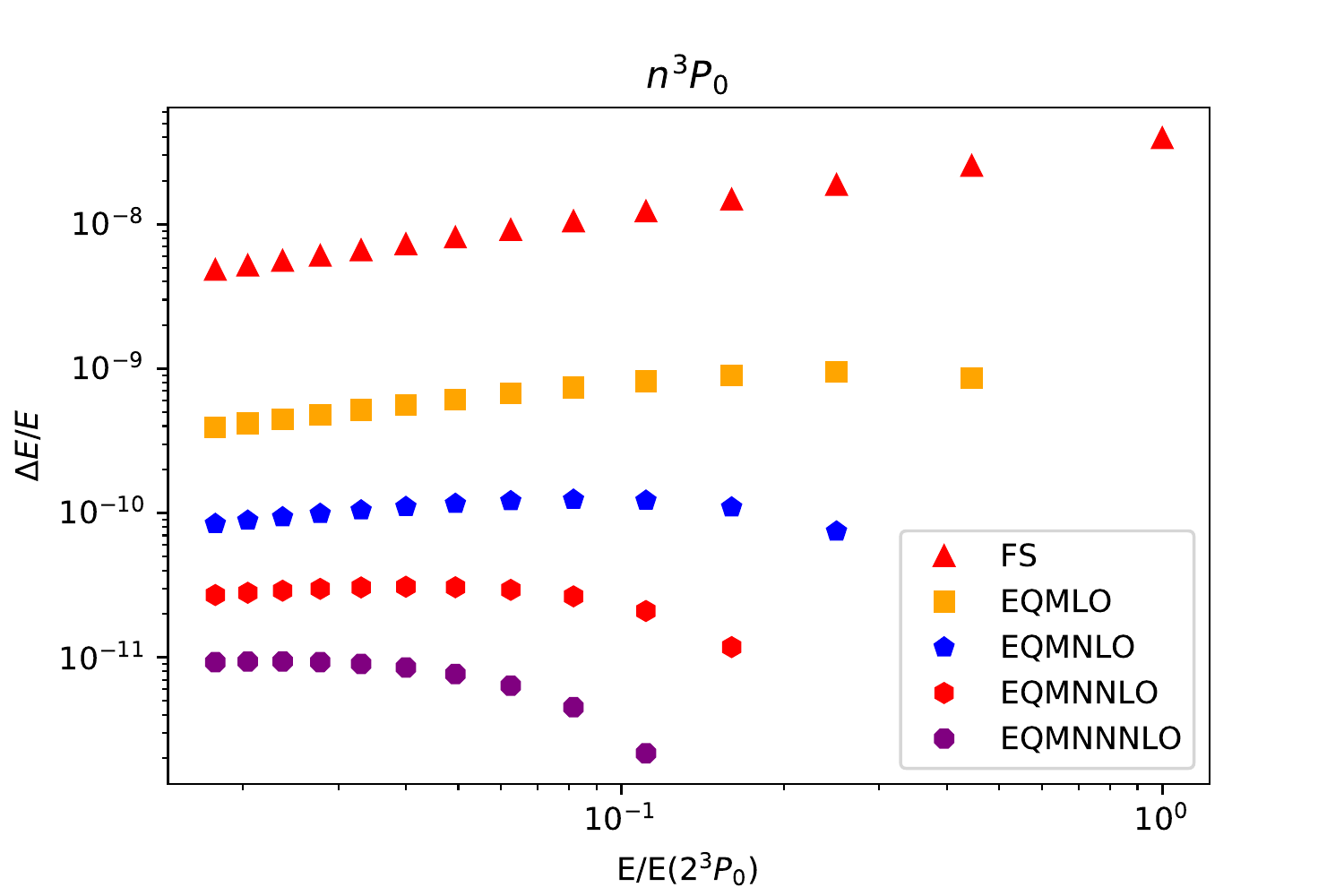}
  \end{center}
  \caption{Relative errors in the predicted $n^3P_0$ state energies, as in Figure \ref{3S1_plot}.}
  \label{3P0_plot}
  \end{figure}

Here we demonstrate how well this effective theory performs, even for moderate values of $n$. To this end, we use the QED predictions of the positronium spectrum, valid up to ${\cal O}\(m\a^6\)$, as a fiducial model \cite{Czarnecki:1999mw}. We use it to fit for the defect parameters and also as a reference for computing the relative error of various models, including the fine structure  corrected (FS) model, which is accurate to order $m\a^4$ (see, e.g., \cite{Cassidy:2018tgq}). We use the effective theory of equation \eqref{the_main_event} for four cases: at lowest order (EQM-LO), we fit only for $\de_{(0)}$; at next-to-lowest order (EQM-NLO)  we fit for both $\de_{(0)}$ and $\l_{(1)}$, and so on. The relative errors of each model's predictions are shown for the $n^3S_1$ and $n^3P_0$ states in Figures \ref{3S1_plot} and \ref{3P0_plot}, respectively.

These results suggest that, given a sufficient number of transition frequency measurements, the predictive power of this theory could exceed the level of precision presently available with QED. It is apparent that within a particular angular momentum channel there will be only a finite number of defect parameters needed to achieve a desired level of accuracy. With a large number of measurements that are sufficiently accurate, a subset of those measurements may be used to fit for the defect parameters which, in turn, may be used to make predictions for the remaining measurements. The comparison of those predictions with their corresponding measurements is one method to check for internal consistency of the entire set of measurements. This approach, or perhaps one utilizing a goodness of fit measure (e.g. \cite{Akaike}) to compare this model with QED, may eventually help to distinguish between a measurement or theoretical error as the explanation of the discrepancy reported in \cite{Gurung:2020hms}.

\begin{figure}[htp]
  \begin{center}
    \includegraphics[width=8.6cm]{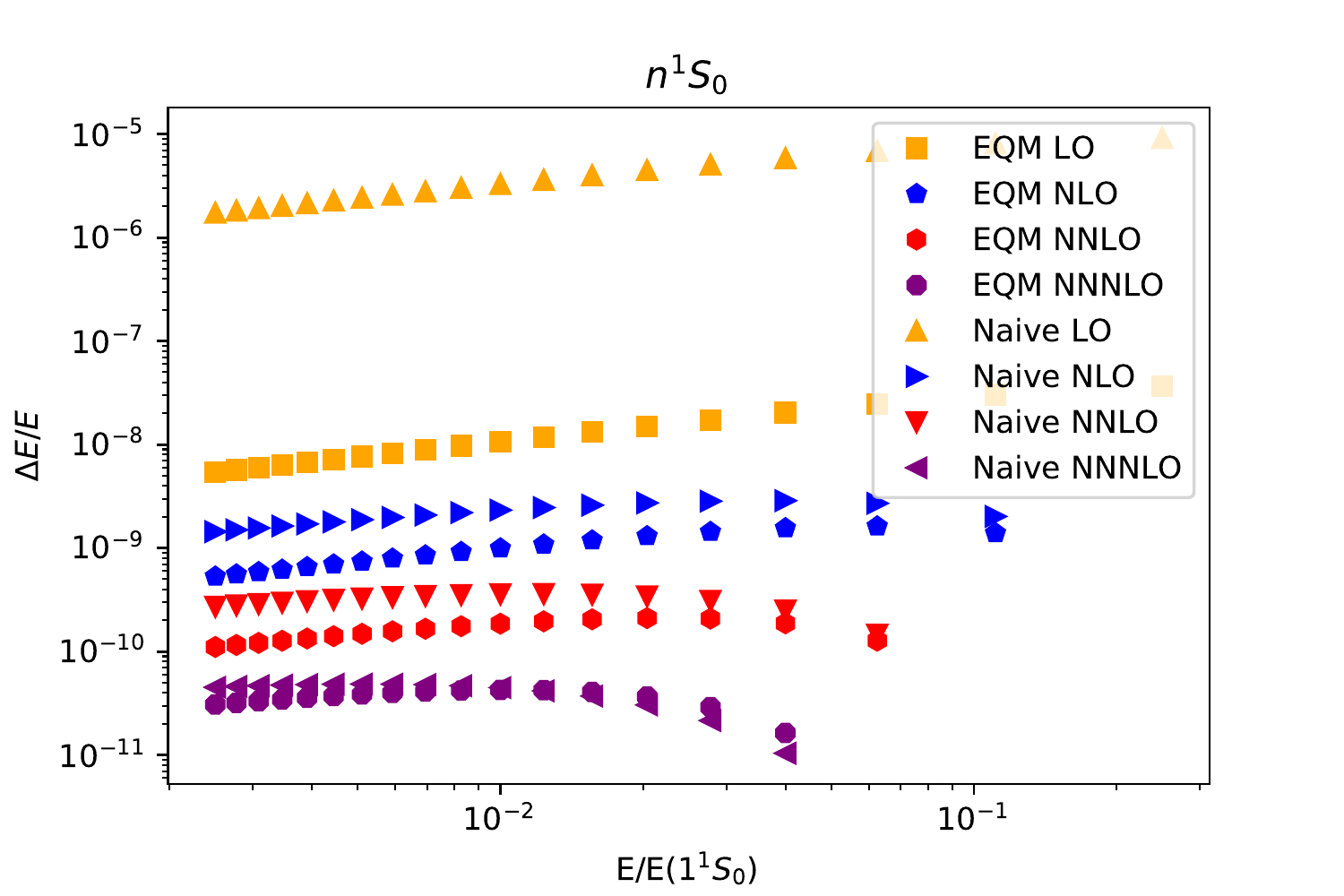}
  \end{center}
  \caption{Relative errors in the predicted $n^1S_0$ state energies using both the effective model, \eqref{the_main_event} and the Naive model, \eqref{naive_ansatz}.}
  \label{1S0_wNaive_plot}
  \end{figure}

It is reasonable to ask if similar results could not be obtained by making a naive ansatz, such as
\begin{equation}\label{naive_ansatz}
E=-\f{1}{4}\f{m\a^2}{n^2} +\f{a_3}{n^3} + \f{a_4}{n^4}  + \f{a_5}{n^5}+\dots \,,
\end{equation}
and fitting for the free parameters, $a_i$. A comparison of the ansatz of \eqref{naive_ansatz} with the effective theory of \eqref{the_main_event} is shown in Figure \ref{1S0_wNaive_plot}, wherein an analysis of the $n^1S_0$ states is given. The Naive LO model uses equation \eqref{naive_ansatz} and has only been fit for $a_3$; the Naive NLO additionally fits for $a_4$, and so on. Although both models provide increased accuracy at higher order, it is clear from Figure \ref{1S0_wNaive_plot} that, order by order, the effective theory of  \eqref{the_main_event} is superior for highly excited states.

\sec{Comparison to QED}\label{Sec:Match_to_QED}

Here we consider why the effective theory, as demonstrated in Figures  \ref{3S1_plot}, \ref{3P0_plot}, and \ref{1S0_wNaive_plot}, agrees so well with the QED predictions. For large $n$, a particular Ps state of a given $\ell, s,$ and $j$ is predicted by QED to have energy levels that may be written as a series\footnote{For example, consider the results of, e.g., \cite{Czarnecki:1999mw}. To show that equation \eqref{omega_expansion} holds, two expansions are required. First, one must utilize an asymptotic expansion for the digamma function for the $S$ states. Second, the so-called Bethe logarithm for all $\ell$ states may be written as an expansion in inverse powers of $n$, starting with $n^{-2}$; however, to the author's knowledge, this has only been demonstrated numerically; see, e.g., \cite{Drake:1990zz, Poquerusse_1981, Jentschura_2005}.} in both $\a$ and $n^{-1}$, i.e.
\begin{equation}\label{omega_expansion}
\o=\sum_{\substack{A=2\\N=2}}^{\infty} C_{A,N} \f{m\a^A}{n^N}\,,
\end{equation}
where the dimensionless coefficients $C_{A,N}$ are ${{\cal O}(1)}$ numbers, some of which may themselves depend on $\a$ through terms like $\ln{\a}$. 
We assume the QED decay rates also take the form
\begin{equation}
\Gamma=\sum_{\substack{A=5\\N=3}}^\infty D_{A,N} \f{m\a^A}{n^N}\,,
\end{equation}
where the $D_{A,N}$ are also assumed to be ${{\cal O}(1)}$ numbers.

Although the value of $\Lambda$ is degenerate with that of the $\l_{\ell s j(i)}$, there are several scales within QED that are natural choices. There is a ``hard", ``soft", and ``ultrasoft" scale, given by $m$, $m\a$,  and $m\a^2$, respectively (see, e.g., \cite{Labelle:1996en} or \cite{Pineda:1998kn}). We make the choice $\Lambda=m\a$, which seems appropriate given that short-distance deviations from the Coulomb potential arise (within QED) from radiative effects involving photons with a characteristic energy $m\a$.

Matching the predictions of the effective theory to QED up to order $m\a^8$, it may be verified that 

\begin{multline}\label{de0_re}
\de_{(0), \text{re}} = -2\(C_{4,3}\a^2+C_{5,3}\a^3 + C_{6,3}\a^4\right.\\
\left. + C_{7,3}\a^5 + C_{8,3}\a^6 \) + {\cal O}\(\a^7\)\,,
\end{multline}
\begin{multline}\label{lambda1_re}
\l_{(1), \text{re}}=8 C_{5,5}\a^2 + \(11C_{4,3}+8C_{6,5}\)\a^3\\
 + \(11C_{5,3}+ 8C_{7,5}\) \a^4 \\
 +\(11C_{6,3}+8C_{8,5}-64C_{4,3}^3\)\a^5 + {\cal O}\(\a^6\)\,,
\end{multline}
\begin{multline}\label{lambda2_re}
\l_{(2), \text{re}}=-32C_{5,7}\a-32C_{6,7}\a^2\\
-\(66C_{5,5}+32C_{7,7}\)\a^3\\
 - \(39C_{4,3}+66C_{6,5} +32C_{8,7}\)\a^4 + {\cal O}\(\a^5\)\,,
\end{multline}
\begin{multline}\label{de0_im}
\de_{(0), \text{im}} = D_{5,3} \a^3 + D_{6,3}\a^4 + D_{7,3}\a^5 \\
+ D_{8,3}\a^6 + {\cal O}\(\a^7\)\,,
\end{multline}
\begin{multline}\label{lambda1_im}
\l_{(1), \text{im}}=- \f{1}{2}\(11D_{5,3}+8D_{7,5}\)\a^4\\
 - \f{1}{2}\(11D_{6,3}+8D_{8,5}\)\a^5 + {\cal O}\(\a^6\)\,,
%\l_{(1), \text{im}}=-4\tc{red}{D_{5,5}}\a - 4\tc{red}{D_{6,5}}\a^2 + \f{1}{2}\(-11D_{5,3}-8D_{7,5}\)\a^3\\
% + \f{1}{2}\(-11D_{6,3}-8D_{8,5}\)\a^4
\end{multline}
and
\begin{equation}\label{lambda2_im}
\l_{(2), \text{im}}= 16D_{7,7}\a^3 + 16D_{8,7}\a^4+{\cal O}\(\a^5\)\,,
\end{equation}
%\l_{(2), \text{im}}=\tc{red}{16D_{6,7}} + \(33\tc{red}{D_{5,5}}+16D_{7,7}\)\a\\
% + \(33\tc{red}{D_{6,5}}+16D_{8,7}\)\a^2+\dots
where we have suppressed the indices $\ell,s,$ and $j$ for simplicity.

However, for the matching indicated in equations \eqref{de0_re} through \eqref{lambda2_im} to be possible, there are consistency relations that must also hold. At the start of this analysis we have only assumed to know the Ps spectrum up to the level of fine-structure corrections, meaning that $C_{4,2}=0$, and all $C_{3,N}=0$, $C_{2,N>2}=0$ and $C_{4,N>4}=0$. For QED matching to be possible, it is necessary that all of the following relations for the $C_{A,N}$ must also hold:
\begin{eqnarray}
C_{5,4} &=& 0\label{C54_consistency}\\ 
C_{5,6} &=& 0\label{C56_consistency}\\ 
C_{5,8} &=& 0\label{C58_consistency}\\ 
C_{6,6} &=& -\f{69}{512}\label{C66_consistency}\\ 
C_{6,8} &=& 0\label{C68_consistency}\\ 
C_{6,4} &=& -3 C_{4,3}^2\label{C64_consistency}\\ 
C_{7,4} &=& -6 C_{4,3} C_{5,3}\\
C_{7,6} &=& -10 C_{4,3} C_{5,5}\\
C_{7,8} &=& -14 C_{4,3} C_{5,7}\\
C_{8,4} &=& \f{3}{4}D_{5,3}^2 - 3C_{5,3}^2 - 6 C_{4,3} C_{6,3}\\
C_{8,6} &=& \!-\f{5}{8} \(11C_{4,3}^2\!+\!16 C_{5,3} C_{5,5} \!+\! 16C_{4,3}C_{6,5}\)\\
C_{8,8} &=&\! \f{1843}{16\,384} \!-\! 7 C_{5,5}^2\!-\!14C_{5,3}C_{5,7}\!-\!14C_{4,3}C_{6,7}\,.
\end{eqnarray}
Similarly, the following consistency relations must hold for the $D_{A,N}$:
\begin{eqnarray}
D_{7,4} &=& -6 C_{4,3} D_{5,3}\label{D74_consistency}\\
D_{7,6} &=& 0\label{D76_consistency}\\%-10 C_{4,3} \tc{red}{D_{5,5}}\\
D_{7,8} &=& 0\label{D78_consistency}\\%-10 C_{4,3} \tc{red}{D_{5,5}}\\
D_{8,4} &=& -6\(C_{5,3}D_{5,3}+C_{4,3}D_{6,3}\)\\
D_{8,6} &=& -10C_{5,5}D_{5,3}\\%-10\(C_{5,5}D_{5,3}+C_{5,3}\tc{red}{D_{5,5}}+C_{4,3}\tc{red}{D_{6,5}}\)\\
D_{8,8} &=& -14C_{5,7}D_{5,3}\,.%-14\(C_{5,5}\tc{red}{D_{5,5}}+C_{4,3}\tc{red}{D_{6,7}}\)\,.
\end{eqnarray}
At present, QED computations for arbitrary $n$ are only available up to ${\cal O}\(m\a^6\)$ in the spectrum; %\flag{what about the $D$'s?} 
therefore only equations \eqref{C54_consistency} through \eqref{C64_consistency} may be verified; however, they are confirmed by all six of the states computed in \cite{Czarnecki:1999mw}, and by nearly all of the (arbitrary) $\ell$ states computed in reference \cite{Zatorski}. The only instance in which \eqref{C64_consistency} appears not to hold is when considering the order $m\a^4/n^3$ term in equation (203) of reference \cite{Zatorski}; however, that term is apparently in error \cite{PrivateConversationAdkins}. Additionally, relations \eqref{D74_consistency}, \eqref{D76_consistency}, and \eqref{D78_consistency} are consistent with the result that those terms are zero \cite{Alekseev:1958,Alekseev:1959}. These relations therefore appear to be robust and provide a useful prediction and consistency check of some of the terms that will appear in QED at order $m\a^7$ and $m\a^8$.

\sec{Conclusions}\label{Sec:Conclusions}

An effective theory of the stationary states of positronium has been developed. This, in what amounts to a relativistic quantum defect theory, can be used to predict and compare spectral measurements without reliance on QED. With future measurements of transition frequencies it may be possible to use this theory to distinguish between a theoretical or experimental error as the explanation of the discrepancy reported in \cite{Gurung:2020hms}. It may have metrological application, as well. With a sufficient number of Ps transition frequency measurements it may be possible to extract the mass of the electron and/or fine structure constant if they are used as fitting parameters.

While the presence of the defect parameters in this theory means that it is less fundamental than QED, it is arguably more robust in the sense that it will accommodate any beyond-QED or BSM physics that are short-ranged, i.e., high-energy, in nature. However, matching this effective theory to the predictions from QED reveals that nontrivial consistency relations exist \emph{within} bound state QED, at least in the context of positronium. These consistency relations have been determined here up to order $m\a^8$, and they will provide an important guidepost for workers performing those calculations.

It remains to be seen how this analysis may be extended to the $m_1\neq m_2$ case, the general class of hydrogenic systems. If this can be achieved, more light may be shed on the nature of bound state QED. Other outstanding questions remain, including if and how it is possible to account for perturbations with this approach, e.g., due to the presence of external fields. Lastly, it is conceivable that some aspects of the analysis presented here could be of use for other two-body systems, such as those involving heavy quarks \cite{Brambilla:2004jw}, for example.

\begin{center}
{\bf Acknowledgements}\\
\end{center}
Many thanks are owed to Bryan Lynn who had suggested that I consider the viability of this effective quantum mechanical method on positronium. I would also like to thank Harsh Mathur for bringing the results of Ref. \cite{Gurung:2020hms} to my attention, and for helpful discussions with Greg Adkins, Yi-Zen Chu, and Matthew Jankowski. A portion of this work was funded by a Charles A. Dana Research Fellowship through Norwich University.

\bibliography{EQM_bib}
\end{document}